\documentclass[twocolumn,secnumarabic,amssymb, nobibnotes, aps, prb]{revtex4-1}
\usepackage{amsmath,amssymb}

\usepackage{graphicx}
\usepackage{wrapfig}
\usepackage{color}
\usepackage{bm}
\usepackage{bbold}

\renewcommand{\i}{\mathrm{i}} 
\renewcommand{\d}{\mathrm{d}} 
\let\daggar=\dag 
\renewcommand{\dag}{^\daggar}
\newcommand{\Tr}{\mathrm{Tr}} 
\newcommand{\up}{_\uparrow} 
\newcommand{\down}{_\downarrow} 

\let\v\undefined 
\DeclareMathAlphabet{\v}{OML}{cmm}{b}{it}

\newcommand{\gv}[1]{\boldsymbol{#1}}
\newcommand{\abs}[1]{\left| #1 \right|} 
\newcommand{\avg}[1]{\left< #1 \right>} 
 
\newcommand{\ket}[1]{\left| #1 \right>} 
\newcommand{\matrixel}[3]{\left< #1 \vphantom{#2#3} \right|#2 \left| #3 \vphantom{#1#2} \right>} 
\newcommand{\tn}[1]{\textnormal{#1}}

\bibpunct{[}{]}{,}{n}{}{} 

\begin{document}

\title{Topological insulator on honeycomb lattices and ribbons without inversion symmetry}

\author{Robert Triebl}\email[Robert Triebl: ]{robert.triebl@tugraz.at}
\author{Markus Aichhorn}

\affiliation{Institute of Theoretical and Computational Physics, NAWI Graz, Graz University of Technology, Petersgasse 16, 8010 Graz, Austria}
\date{\today}

\begin{abstract}
 We study the Kane-Mele-Hubbard model with an additional inversion-symmetry-breaking term. Using the topological Hamiltonian approach, we calculate the $\mathbb{Z}_2$ invariant of the system as function of spin-orbit coupling, Hubbard interaction $U$, and inversion-symmetry-breaking on-site potential. The phase diagram calculated in that way shows that, on the one hand, a large term of the latter kind destroys the topological non-trivial state. On the other hand, however, this inversion-symmetry-breaking field can enhance the topological state, since for moderate values the transition from the non-trivial topological to the trivial Mott insulator is pushed to larger values of interaction $U$. This feature of an enhanced topological state is also found on honeycomb ribbons. With inversion symmetry, the edge of the zigzag ribbon is magnetic for any value of $U$. This magnetic moment destroys the gapless edge mode. Lifting inversion symmetry allows for a finite region in interaction strength $U$ below which gapless edge modes exist.
 \end{abstract}

\maketitle

\section{Introduction}

Since topological insulators have been theoretically predicted 10 years ago~\cite{kanemele1, kanemele2}, the understanding of topological phases has progressed enormously. Topological Hamiltonians are classified by the tenfold way~\cite{kitaev_periodic, schnyder1, schnyder2}, various experiments have been performed showing the practical relevance of the theoretical considerations~\cite{konig_quantum_2007,roth_nonlocal_2009, hsieh_topological_2008, zhang_topological_2009, hsieh_observation_2009, nishide_direct_2010,xia_observation_2009, hasan}, and several groups already succeeded in a next step which is predicting and realizing Weyl semimetals~\cite{fang_anomalous_2003, wan_topological_2011, xu_chern_2011,  huang_weyl_2015, yang_weyl_2015, xu_discovery_2015, weng_weyl_2015, lv_experimental_2015}.

However, the influence of interactions onto the topological classification is still not fully understood. Just recently, new phase transitions in strongly correlated topological insulators have been reported~\cite{amaricci1, amaricci2}. The most used quantity to characterize topological order, namely the $\mathbb{Z}_2$ invariant introduced by Fu, Kane, and Mele~\cite{kanemele1,fu1,fu2, fu_topological_2007}, relies on defined Bloch bands and is thus not directly applicable for interacting systems. A generalization is possible using the so-called topological Hamiltonian~\cite{wang_inversion, wang_X, wang_topH}, an artificially noninteracting system determined by the Green's function. 

The Kane-Mele-Hubbard (KMH) model~\cite{kanemele1,kanemele2,rachel} combines a topological model Hamiltonian with strong interactions and is therefore frequently used to explore correlation effects in topological insulators~\cite{rachel, lee, hohenadler1, hohenadler2, assaad, hohenadler,hung,  yu, budich, laubach, wu, grandi, chen, lai, miyakoshi}. Within the framework of the topological Hamiltonian, the calculation of the $\mathbb{Z}_2$ invariant is straight forward as long as inversion symmetry is obeyed, since only the time-reversal-invariant momenta (TRIMs) have to be considered~\cite{fu2, wang_inversion}. In case of the bare KMH model, it can thus be used since inversion symmetry is respected~\cite{budich, laubach, hohenadler}. 

Determining the topological phase becomes more difficult if an inversion-symmetry-breaking term such as a staggered on-site potential~\cite{kanemele1, lai}, a Rashba coupling~\cite{kanemele1, laubach}, or site-dependent hoppings~\cite{chen,hung} are included. A possibility to analyze topological phases is to calculate the spin Chern number~$C_S$~\cite{sheng2, hung, chen, lai, yoshida, miyakoshi}. This approach requires spin to be a good quantum number and has the drawback that due to numerical artifacts a good quantization of $C_S$ is not given close to phase transitions. Another approach is to look directly for gapless edge states and use bulk-boundary correspondence~\cite{yu, laubach, wu, grandi}. 

In this paper, we calculate the $\mathbb{Z}_2$ invariant of the KMH model with an inversion-symmetry-breaking on-site potential by combining the topological Hamiltonian with a method introduced by Soluyanov and Vanderbilt~\cite{solvan, solvan2} that is based on maximally localized Wannier charge centers~\cite{marzari}. This enables a precise calculation of invariants without restricting the systems to certain symmetries. Furthermore, we investigate bulk-boundary correspondence by calculating the spectral functions of a zigzag ribbon. We show that bulk-boundary correspondence has to be treated with care in strongly interacting systems since time-reversal symmetry might be lifted locally at the edges due to spontaneous symmetry breaking. 
The Green's functions in our approach are obtained by a two-site dynamical impurity approximation~\cite{potthoff1, potthoff2, vca, aichhorn, dahnken}.

\section{Model and Methods}
\subsection{Kane-Mele-Hubbard model}
The Kane-Mele-Hubbard Hamiltonian is used exemplary since it is a toy model for strongly correlated topological insulators. The noninteracting part as proposed by Kane and Mele~\cite{kanemele1,kanemele2} is given by
\begin{eqnarray}\label{eq:KM}
 H_\tn{KM}&=&-t \sum_{\langle i,j\rangle} c_i\dag c_j + \i \lambda_\tn{SO} \sum_{\langle\langle i,j\rangle\rangle} \nu_{ij} c_i\dag \sigma^z c_j \nonumber \\
 &&+ \lambda_\nu \sum_i \xi_i c_i\dag c_i
\end{eqnarray}
on a honeycomb lattice, where $c_i\dag$ is the creation operator of a spinor $\left(c_{i\uparrow}\dag, c_{i\downarrow}\dag\right)$, $\langle \cdot \rangle$ denotes nearest neighbors, and $\langle \langle \cdot \rangle \rangle$ next-nearest neighbors. The first term is a tight-binding nearest-neighbor hopping term, which is commonly used to model the Dirac cones of graphene up to first order. The second is the intrinsic spin-orbit coupling, leading to the quantum spin Hall topological insulating state as it opens a gap~\cite{kanemele1,kanemele2, sheng2}. The third term is a staggered on-site potential, where $\xi_i= 1$ if site $i$ belongs to sublattice $A$ of the honeycomb lattice, and $-1$ if it belongs to sublattice $B$. This distinction between the two sublattices breaks inversion symmetry and has a crucial influence on topology: the KM model with a sublattice potential is a topological insulator for any finite $\lambda_\tn{SO}$, as long as $\abs{\lambda_\nu}<3\sqrt{3}\lambda_\tn{SO}$. The gap closes for $\abs{\lambda_\nu}=3\sqrt{3}\lambda_\tn{SO}$ and reopens for $\abs{\lambda_\nu}>3\sqrt{3}\lambda_\tn{SO}$, but the topology becomes trivial in that case. 

Interaction effects can be introduced by a Hubbard interaction $U n_{\uparrow} n_{\downarrow}$ on each site, leading to the Kane-Mele-Hubbard Hamiltonian~\cite{rachel}
\begin{equation}\label{eq:hubbard}
H_\tn{KMH} = H_\tn{KM}+U\sum_i n_{i \uparrow} n_{i \downarrow}.
\end{equation}
Throughout the paper, the energy scale is defined by ${t\equiv 1}$, and the length scale by the lattice parameter ${a\equiv 1}$.

\subsection{Calculation of topological invariants}

Topological systems are classified by their dimension and symmetries, as summarized in the periodic table of topological matter~\cite{kitaev_periodic, schnyder1, schnyder2}. The important symmetry in case of the KM model is time reversal, leading to the topological class AII, specified by a $\mathbb{Z}_2$ invariant $\nu$. A possibility to define this invariant is via time-reversal polarizations of a one-dimensional system that depends on an additional pumping parameter, as introduced by Fu and Kane~\cite{fu1}. In case of a noninteracting two-dimensional system, this definition is applicable if $k_y$ is considered as the pumping parameter. For the actual calculation of $\nu$, inversion-symmetric and non-inversion-symmetric systems are treated differently, as discussed in the following.

If inversion symmetry is present, the four TRIMs $\gv{\Gamma}_i$ contain the whole topological information. The $\mathbb{Z}_2$ invariant $\nu$ can be obtained by 
\begin{equation}\label{eq:nu_inv}
  (-1)^\nu=\prod_{i=1}^4 \delta_i \quad\textnormal{with} \quad \delta_i=\prod_{n=1}^N \xi_n(\gv{\Gamma}_i),
\end{equation}
where $\xi_n(\gv{\Gamma}_i)$ is the eigenvalue of the parity operator at momentum $\v{k}=\gv{\Gamma}_i$ of Kramer's pair $n$~\cite{fu2}.

\begin{figure}
 \centering
  \includegraphics[width=0.37\textwidth]{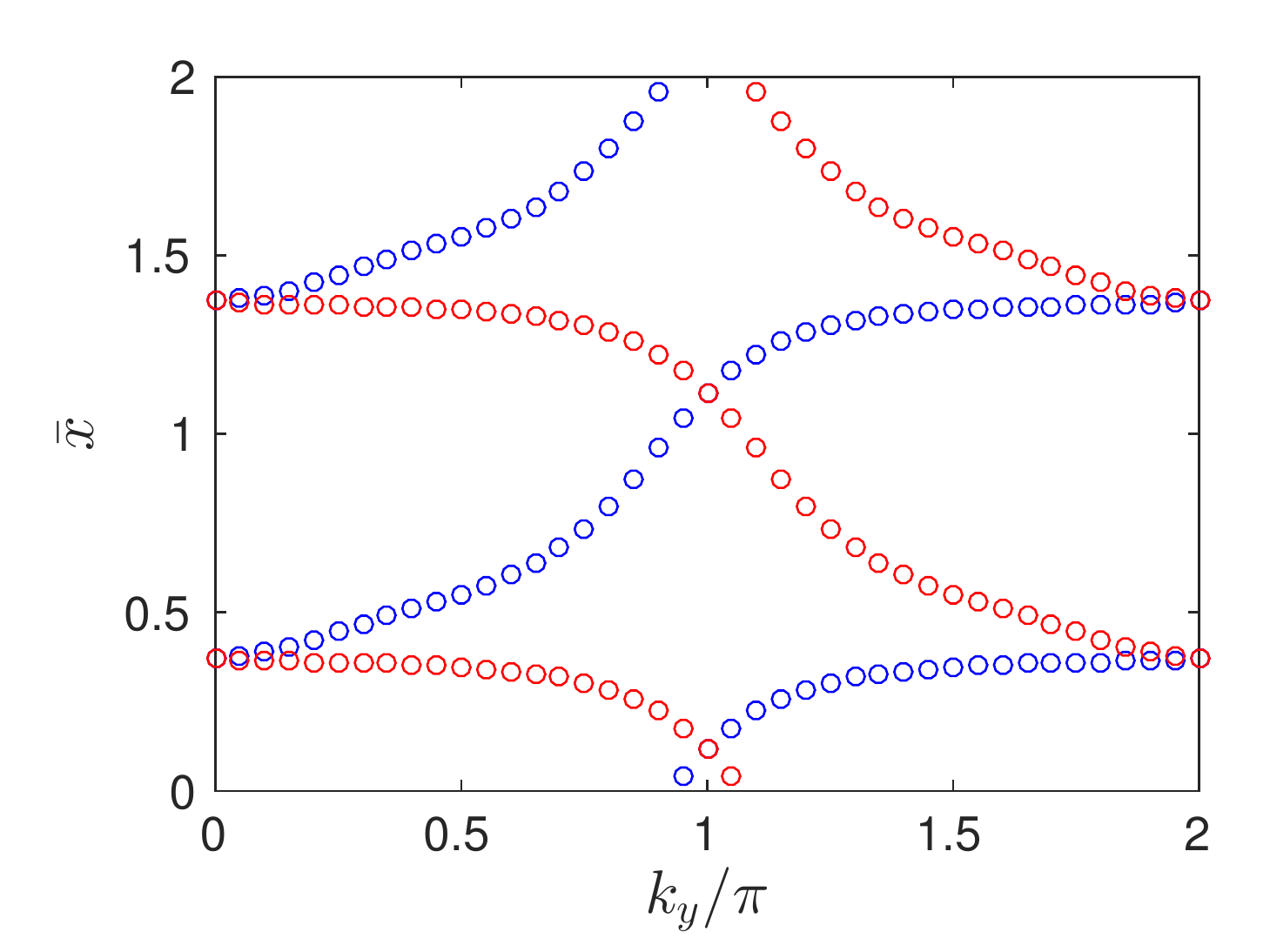} 
  \includegraphics[width=0.37\textwidth]{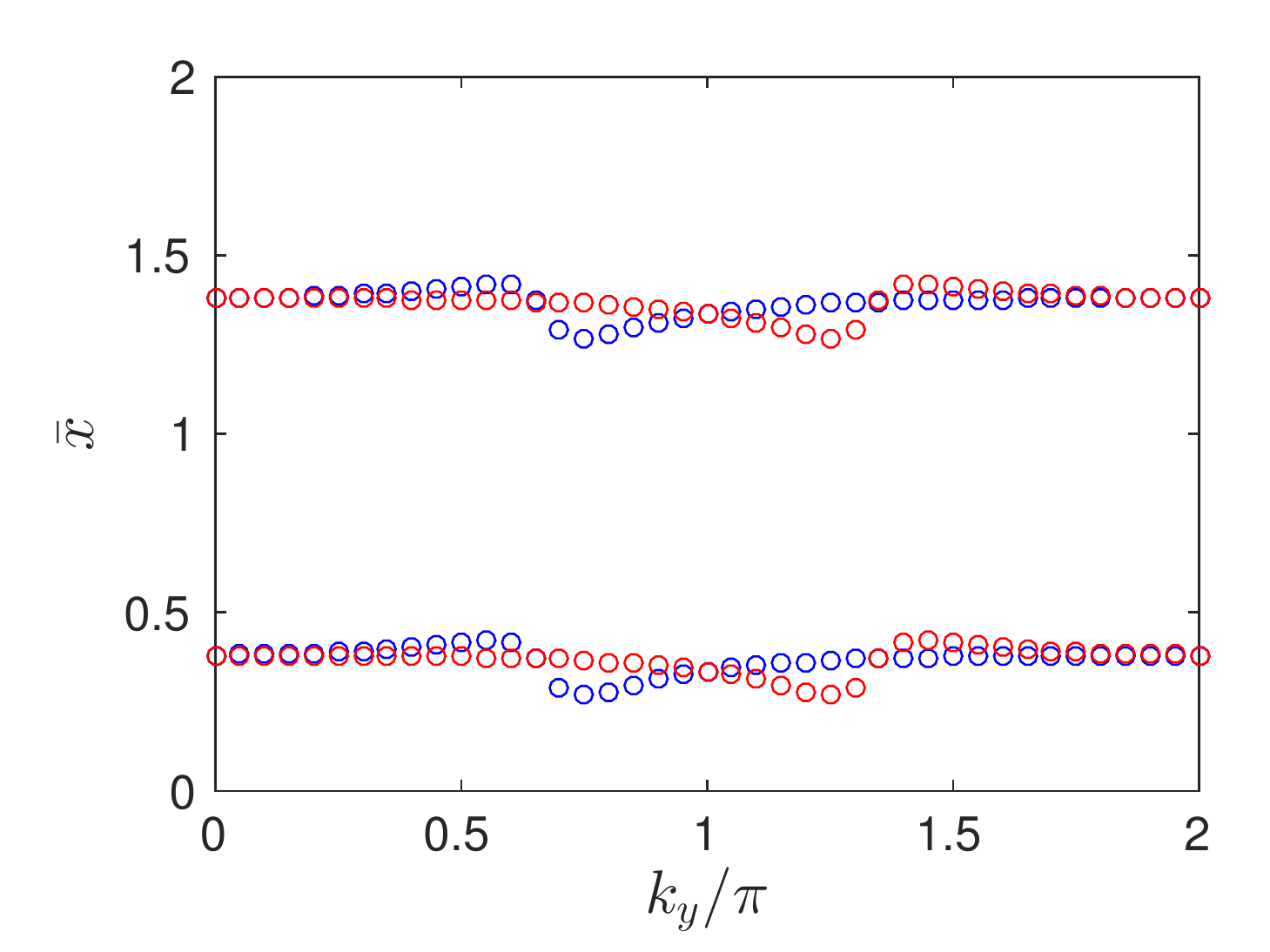}
 \caption{Examples of WCCs according to the noninteracting KM model ($\lambda_\tn{SO} = 0.5$) with nontrivial (top, $\lambda_\nu = 2.5 \lambda_\tn{SO}$) and trivial (bottom, $\lambda_\nu = 5.5 \lambda_\tn{SO}$) topology. Here, two unit cells are shown along the abscissa so that at least one WCC is continuously displayed. Thus, four instead of two WCCs are visible. $S_z$ is conserved, so the WCCs can be separated in spin up (blue) and spin down (red). \label{fig:1}}
\end{figure}

If, on the other hand, inversion symmetry is broken, one needs information on how the Bloch states evolve continuously between the TRIMs. Soluyanov and Vanderbilt suggested~\cite{solvan, solvan2} to use hybrid Wannier functions  
\begin{equation}\label{eq:hwf}
 \ket{R_xk_yn}=\frac{1}{2\pi}\int_{-\pi}^\pi\d k_x\;e^{-\i R_x k_x} \ket{\psi_{n\v{k}}},
\end{equation} 
which are maximally localized~\cite{marzari}. The topology is determined by tracking the maximally localized Wannier charge centers (WCCs) assigned to the occupied bands along the pumping parameter $k_y$, which are given by  $\bar{x}_n(k_y) = \matrixel{0k_yn}{x}{0k_yn}$. This function is defined modulo a lattice constant that is chosen to be 1, so $\bar{x}(k_y)$ has a periodicity of $2\pi$ in $k_y$, and a period of $1$ along $\bar{x}$. The KM model has only two occupied bands which form a Kramers pair because of time-reversal invariance, so Kramer's degeneracy enforces the two WCCs to be equal at $k_y=0$ and $\pi$. Tracking the WCCs continuously from $k_y=0$ to $2\pi$, the system is trivial if the very same WCCs intersect at both points, and nontrivial if there is a shift which is a multiple of the lattice constant. Examples are given in Fig. \ref{fig:1}. If the spin in $z$ direction $S_z$ is conserved, the continuous tracking is straight forward since each WCC can be assigned to a certain spin. If no conserved quantity helps identifying the respective WCCs, the two cannot be sorted and some more advanced method has to be applied, as for example tracking the difference of the WCCs~\cite{solvan}.

We now turn to the determination of topological states for a system with electron-electron interactions. Here, topological invariants cannot be defined as described above since one-electron Bloch functions are not eigenstates. A more general definition of the first Chern number uses Green's functions~\cite{gurarie,volovik,niu},
  \begin{equation}\label{eq:cgreen}
  C=\frac{\epsilon^{\mu\nu\rho}}{24\pi^2}\int \d k_0 \int \d^2 k \;
  \Tr\left[G\partial_\mu G^{-1}G \partial_\nu G^{-1}G \partial_\rho G^{-1}\right]
  \end{equation}
with $k_0 = \i \omega$, which gives the integer coefficient of the
quantum Hall effect of a two dimensional system. If spin is a good
quantum number, the Chern invariant can be evaluated separately for
each spin. $C\up$ is then evaluated from the spin up block of the Green's function, $C\down$ from the spin down block. This leads to a quantized spin Chern number $C_S = (C\up -C\down) / 2$, which is integer for time-reversal invariant
Hamiltonians. Modulo 2, this quantity can be used as a $\mathbb{Z}_2$ invariant. In the general case, a $\mathbb{Z}_2$ invariant $\nu$ is obtained from a dimensional
reduction of the second Chern number 
 \begin{eqnarray}
  C_2 &=& \frac{\epsilon^{\mu \nu \rho \sigma \tau} }{480\pi^3} \int \d k_0 \int \d^4 k \;
  \Tr\left[G\partial_\mu G^{-1}G \partial_\nu G^{-1} \right.\nonumber\\
  && \times \left. G \partial_\rho G^{-1}G \partial_\sigma G^{-1}G \partial_\tau G^{-1} \right], \label{eq:cgreen2} 
 \end{eqnarray}
 which describes the response of a four dimensional insulator~\cite{qi_TFT, wang_PRL}. 
Starting from definitions (\ref{eq:cgreen}) and (\ref{eq:cgreen2}), Wang \textit{et al.} showed that the topological information is already captured in the Green's function at zero frequency~\cite{wang_inversion, wang_X, wang_topH}. They conclude that minus the inverse Green's function at zero frequency can be considered as the Bloch Hamiltonian of an artificial noninteracting system which has the same Chern invariant and the same $\mathbb{Z}_2$ invariant as the interacting one, as long as they are continuously connected. Thus, this Bloch Hamiltonian is called topological Hamiltonian~\cite{wang_topH}
\begin{equation}\label{eq:Htop}
 H_t(\v{k})=-G^{-1}(\omega=0,\v{k})
\end{equation}
of the interacting system. A main consequence is that methods devised for noninteracting
Hamiltonians are sufficient to calculate topological numbers related
to the more complicated integrals (\ref{eq:cgreen}) and
(\ref{eq:cgreen2}), as for example $C_S$ and $\nu$. A direct
evaluation of (\ref{eq:cgreen}) or (\ref{eq:cgreen2}) is therefore not necessary.

If the system obeys inversion symmetry, $G^{-1}(\omega,\v{k})$ commutes at the TRIMs $\v{k}=\gv{\Gamma}_i$ with the parity transformation matrix $P$, and as a consequence, ther are simultaneous eigenstates $\ket{\alpha(\omega,\gv{\Gamma}_i)}$ of $G^{-1}$ and $P$:
\begin{equation}
P\ket{\alpha(\omega=0,\gv{\Gamma}_i)}=\eta_\alpha\ket{\alpha(\omega=0,\gv{\Gamma}_i)}.
\end{equation}
The topological invariant $\nu$ can be calculated from these eigenvalues through~\cite{wang_inversion}
\begin{equation}\label{eq:nu_inv_int}
(-1)^\nu=\prod_\tn{R zeros}\eta_\alpha^{1/2}.
\end{equation}
Here, the convention $(-1)^{1/2}=+\i$ is used. In the noninteracting case, this equation reduces to the Fu-Kane formula (\ref{eq:nu_inv})~\cite{wang_inversion}. 

The direct evaluation of topological invariants through Eq.~(\ref{eq:nu_inv_int}) became already a standard procedure in case of interacting systems with inversion symmetry~\cite{budich,laubach,hohenadler, amaricci1, amaricci2}.
In this work, we are interested in the topological invariants of an
interacting system \textit{without inversion symmetry}, where
Eq.~(\ref{eq:nu_inv_int}) cannot be applied. For this case we propose
to use a combination of the topological Hamiltonian with the
Soluyanov-Vanderbilt method of WCCs as described in the beginning of
this section. In practice, we first calculate the Green's function at
zero frequency using a dynamical impurity approximation as explained
in the next section. The obtained topological Hamiltonian can then be
used just like a Bloch Hamiltonian to determine the $\mathbb{Z}_2$
invariant. This in turn is done with Wannier charge centers as
proposed by Soluyanov and Vanderbilt~\cite{solvan}, just using the
eigenstates of the topological Hamiltonian
$\ket{\alpha(\omega=0,\v{k})}$ instead of the Bloch functions
$\ket{\psi_{n\v{k}}}$ of the noninteracting case.

\subsection{Variational cluster approach}
As described in the previous section, the one-electron Green's function is needed to determine the topological Hamiltonian. Since an exact solution of the full many-body problem is not possible, an approximative method has to be chosen. Here we apply the Variational cluster approach (VCA)~\cite{potthoff1,vca}, because the Kane-Mele-Hubbard model is known to have an antiferromagnetic moment~\cite{rachel, hohenadler1, hohenadler2, hohenadler, lee,laubach, budich} which can efficiently be treated by the VCA with symmetry-breaking Weiss fields~\cite{dahnken,aichhorn}.

The VCA is based on the self-energy functional approach, which uses the fact that the grand potential of an arbitrary interacting system  $H=H_0(\v{t})+H_1(\v{U})$ has to be a stationary point of the self-energy functional 
\begin{equation}
\Omega_\v{t}[\gv{\Sigma}]\equiv \Tr \log \left(-(\v{G}_0^{-1}-\gv{\Sigma})^{-1}\right)+F[\gv{\Sigma}],
\end{equation} 
where $F[\gv{\Sigma}]$ denotes the Legendre transform of the Luttinger-Ward functional $\Phi[\v{G}]$~\cite{potthoff1,luttinger}. The approximation of this method is to restrict the space of self-energies $\gv{\Sigma}$. This subset $\mathcal{S}$ of self-energies is spanned by all $\gv{\Sigma}(\v{t}')$ that are the exact self-energies of a so-called reference system $H'=H_0(\v{t}')+H_1(\v{U})$. The interaction parameters $\v{U}$ are the same as in the original system, but $H$ and $H'$ can differ in the one-particle parameters. The one-particle parameters $\v{t}'$ of the reference system $H'$ are chosen such that the self-energy of the reference system can be calculated exactly. To obtain the approximative physical self-energy  $\gv{\Sigma} \in \mathcal{S}$, a stationary point of $\Omega_\v{t}[\gv{\Sigma(\v{t}')}]$ has to be found as $\v{t}'$ is varied. The parametrized functional can be reduced to 
\begin{eqnarray}\label{eq:vca_gp}
\Omega_\v{t}[\gv{\Sigma}(\v{t}')]&=& \,\Omega'(\v{t}')+\Tr \log \left(-\left(\v{G}_0^{-1}(\v{t})-\gv{\Sigma}(\v{t}')\right)^{-1}\right)\nonumber\\
&&-\Tr \log \left(-\left(\v{G}_0^{-1}(\v{t'})-\gv{\Sigma}(\v{t}')\right)^{-1}\right)
\end{eqnarray}
and can thus be calculated if the Green's function of the reference system is known. Quite generally, reference systems in the VCA are clusters of finite size, which can be treated by exact diagonalization techniques~\cite{potthoff1,vca,aichhorn}.

In case of the KMH model, several cluster sizes have already been analyzed~\cite{yu, budich, laubach}.
However, the tiling of the lattice into clusters of finite sizes breaks artificially some symmetries, which can change the topological phase diagram~\cite{varney}.
That is why we choose as a reference system for VCA single-site clusters, which are coupled to one additional bath site by a hopping $V$.
This rather simple approach, called two-site dynamical impurity approximation (DIA)~\cite{potthoff2}, has two advantages. First, despite its simplicity, it gives accurate results for the transition towards an antiferromagnetic insulator for two-dimensional Hubbard models~\cite{potthoff2}. Second, which is even more important, the lattice symmetries are trivially satisfied. A drawback of this method is the locality of the self-energy. We will show below, however, that for known cases we get very good agreement with existing results obtained by numerically much more expensive methods.

\begin{figure}
 \centering
   \includegraphics[width=0.43\textwidth]{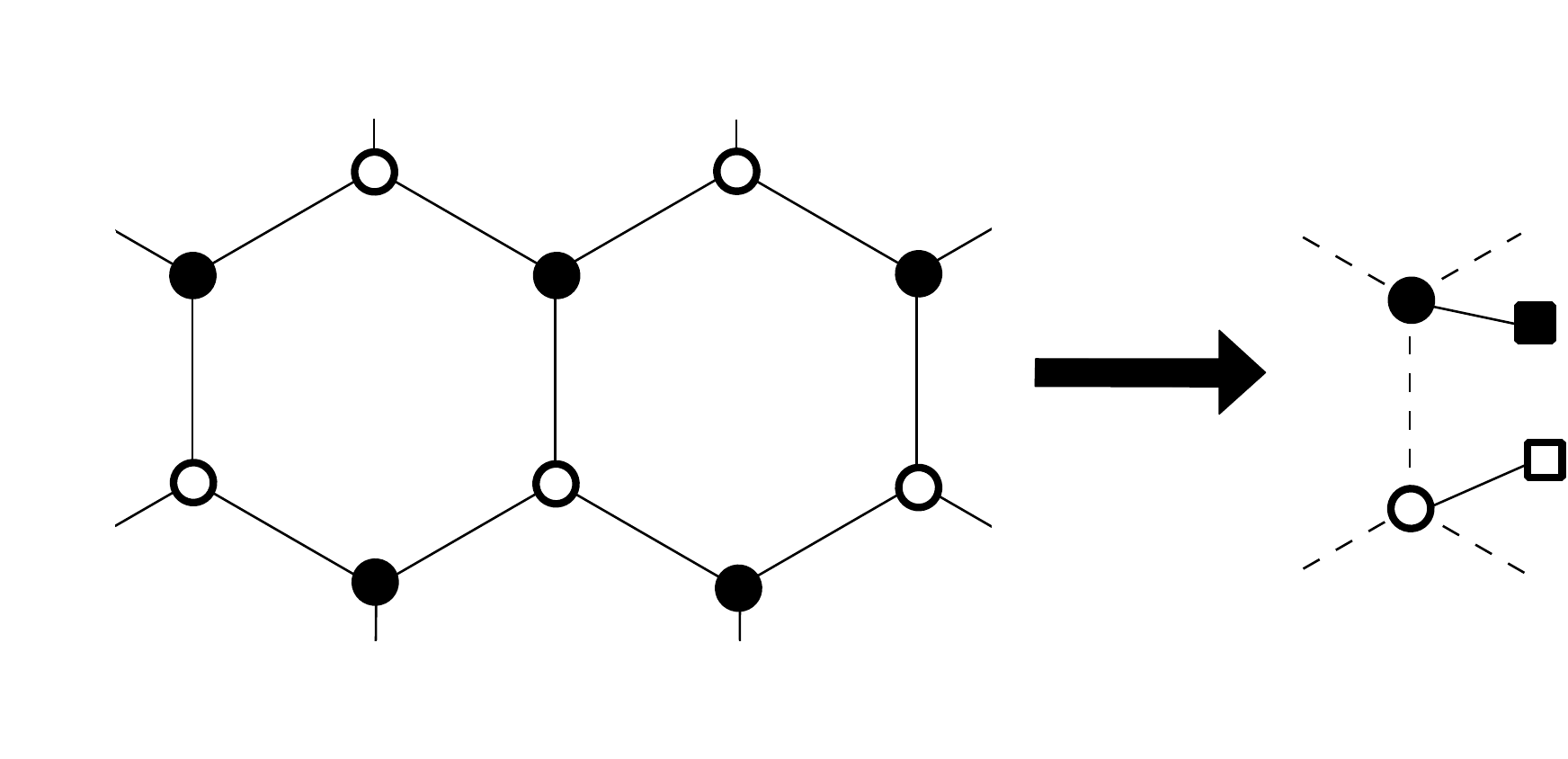}
 \caption{The left plot shows the full system, the right the reference system. Full symbols denote sublattice $A$, open symbols sublattice $B$. The bath sites (squares) are characterized only by an on-site energy. The impurity sites (circles), on which the Hubbard $U$ is acting, can additionally carry the symmetry-breaking Weiss fields. \label{fig:2}}
\end{figure}

Since the honeycomb lattice has two distinct sites, the unit cell is tiled by two clusters, which are coupled by the noninteracting part of the Hamiltonian, as shown in Fig. \ref{fig:2}. On-site energies on both impurity and bath site, as well as the connecting hopping between them, give in total three variational parameters per cluster. However, in the inversion-symmetric case ($\lambda_\nu=0$), the on-site energies are fixed by particle-hole symmetry and only one parameter remains.

In order to capture symmetry breaking necessary for the emerging antiferromagnetic moment, a Weiss field 
\begin{equation}\label{eq:vca_weiss}
H_\tn{AF}=\sum_i c_i\dag \left( \v{h}_i \cdot \gv{\sigma}\right) c_i
\end{equation}
has to be added~\cite{dahnken}. Without any symmetry considerations, these fields on both $A$ and $B$ sites give in total 6 variational parameters. Due to the inversion-symmetry breaking on-site potential $\lambda_\nu$, a second Weiss field 
\begin{equation}\label{eq:vca_weiss2}
H_\Delta=\Delta \sum_i \xi_i c_i\dag c_i
\end{equation}
is used to enable unequal electron densities on the two sublattices. As in Eq. (\ref{eq:KM}), $\xi_i=\pm1$, depending on the sublattice. This Weiss field is basically a renormalisation of $\lambda_\nu$ in the reference system, which is caused by the interplay of the sublattice potential and Hubbard interaction.

\begin{figure}
 \centering
   \includegraphics[width=0.45\textwidth]{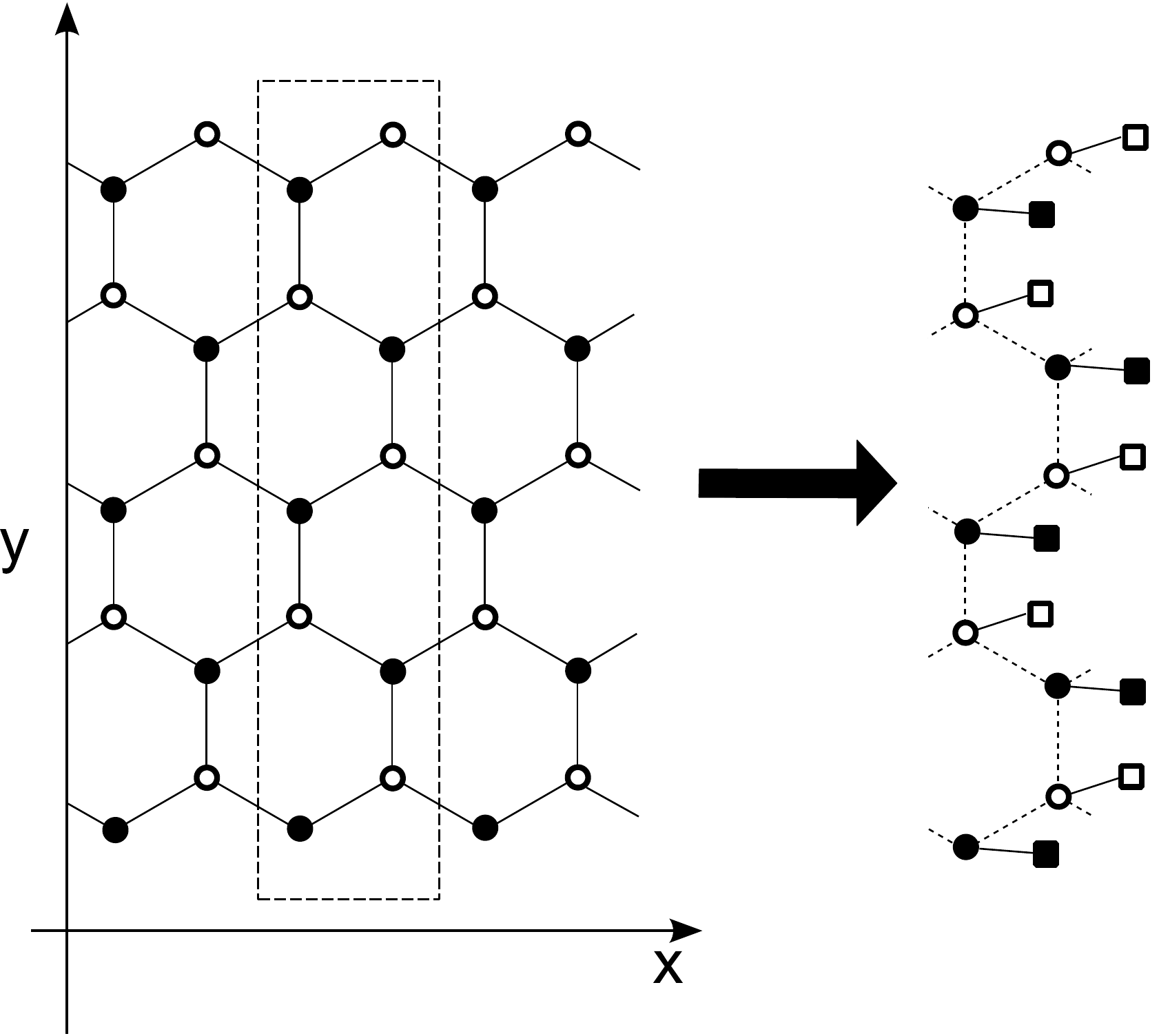}
 \caption{Unit cell of the zigzag ribbon and the according reference system. The respective two-site clusters are identical, except for a different AF Weiss field. \label{fig:3}}
\end{figure}

The method described so far considers bulk properties. Introducing an edge destroys translational symmetry and influences therefore local magnetization. As known from field theoretical investigations, mean-field approximation gives a finite magnetization on the zigzag edge for every finite interaction strength~\cite{lee}. This could lead to a breakdown of the bulk-boundary correspondence and may cause problems for calculating topological invariants using the existence of gapless edge states as a proof for nontrivial topology, which has so far been used in some cases of interacting systems without inversion symmetry~\cite{yu, laubach,grandi}. Vice versa, a nontrivial topological invariant in the bulk may not result in gapless edge states due to locally broken time-reversal symmetry caused by spontaneous symmetry breaking. Therefore, we additionally implemented the DIA on the zigzag ribbon in order to compare the topological invariants defined by the bulk Green's function to the existence of gapless edge states. The ribbon is translationally invariant in the $x$ direction, whereas the sites along the width of the ribbon are distinct. If a unit cell contains $N$ pairs of $A$ and $B$ sites, 2$N$ clusters containing each a bath and an impurity site have to be solved and effectively coupled by the noninteracting part of the Hamiltonian (see Fig. \ref{fig:3}). In order to keep the number of parameters manageable, the on-site energies and hybridisations are chosen to be constant along the ribbon. To allow for edge magnetization, the antiferromagnetic Weiss fields for each pair of sites $A$ and $B$ is varied independently, only assuming a mirror symmetry $y \mapsto -y$.

\section{Results}

\subsection{Bulk}\label{bulk}
\begin{figure}
 \centering
  \includegraphics[width=0.47\textwidth]{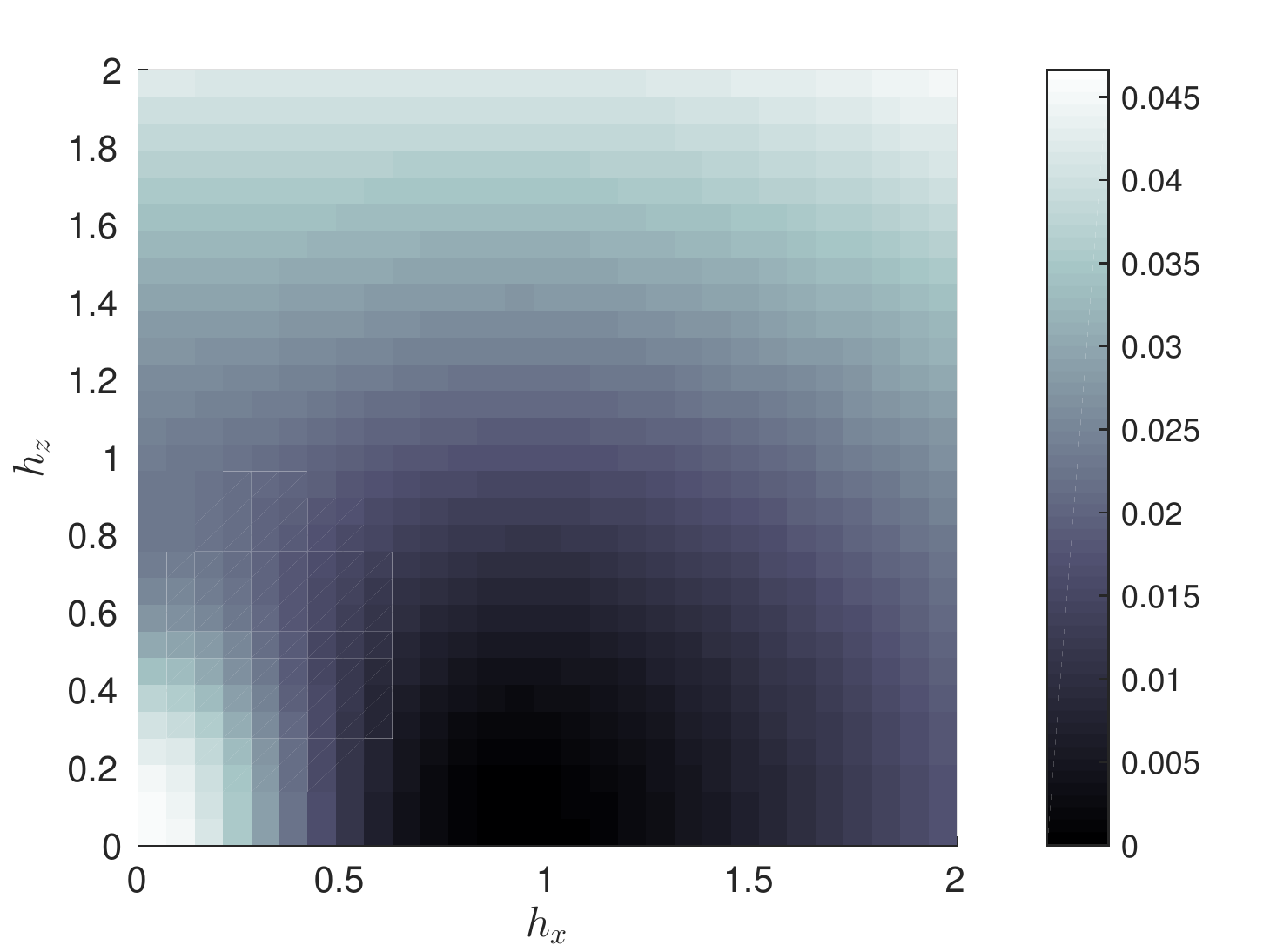}
 \caption{Self-energy functional as a function of the antiferromagnetic Weiss fields $h_z$ and $h_x$ for $\lambda_\tn{SO}=0.1$, $\lambda_\nu=0$ and $U=5$. The hybridisation of the bath sites has been optimized for each grid point individually. The global minimum around $h_x\approx 1$ and $h_z=0$ can clearly be seen. \label{fig:4}}
\end{figure}

As mentioned in the methods section, the hopping to the bath sites, the magnetic Weiss fields, and the sublattice potential Weiss field have to be determined in the VCA. For all stationary points, the ferromagnetic part of the Weiss field vanishes, hence only an antiferromagnetic ordering $\v{h}_A = -\v{h}_B$ is possible. Without spin-orbit coupling, the system has full SU(2) symmetry, so only the absolute value of the Weiss field has to be determined. When spin-orbit coupling is included, only the $xy$-plane is still degenerate, but the degeneracy of the $z$ direction is lifted. This means that we have to deal with two antiferromagnetic Weiss fields, $h_z$ and $h_{x}$. 
To analyze the direction of the antiferromagnetic moment, we calculate a two-dimensional surface of the self-energy functional $\Omega(h_z,h_x)$, where all other variational parameters are optimized for each set of variables $(h_z, h_x)$. The stationary points, i.e. extrema and saddle points, of this two-dimensional surfaces are physical solutions, where the stable solution is the one with lowest potential $\Omega$. 
Fig.~\ref{fig:4} shows the value of the self-energy functional as a function of both in-plane and out-of-plane AF symmetry-breaking field. Depending on the KMH model parameters, up to three different stationary points exist: A saddle point of $\Omega$ if $\v{h}$ points in $z$ direction; a minimum if it is in the $xy$ plane; the nonmagnetic solution, which can be both maximum or minimum, depending on the parameters. This is consistent with the results of other cluster geometries~\cite{yu, laubach}. The local minimum $\v{h} \parallel \hat{z}$ is never the physically realized solution with the lowest grand potential $\Omega$ for all sets of parameters considered here. Hence, only one variational quantity is needed for the AF Weiss field, namely the in-plane antiferromagnetic component. As mentioned above, the on-site energy levels of both impurity and bath are fixed by particle hole symmetry and the given chemical potential. Therefore, in total three cluster parameters have to be optimized: The hopping $V$ between impurity and bath, the in-plane antiferromagnetic Weiss field $h_x$, and the potential difference between the two sublattices $\Delta$. 

Directly from the two-site DIA one can distinguish two phases, the antiferromagnetic insulator for large $U$ and the nonmagnetic insulator for small $U$. The system reduces to the ordinary Hubbard model on the honeycomb lattice if $\lambda_\tn{SO}=0$ and $\lambda_\nu=0$. In this case, the magnetization direction is not important since $SU(2)$ symmetry is not broken. The mean-field critical interaction is $U_c = 2.23$~\cite{rachel, sorella}, which is lower as compared to more accurate methods. Quantum Monte Carlo simulations show that it is actually slightly above 4~\cite{sorella, hohenadler1, hohenadler2, assaad, hohenadler}. The two-site DIA considered in this work is expected to give similar results as other variational methods. VCA gives critical interactions between 2.4 and 4, depending on the cluster geometries~\cite{yu, budich,laubach}, which coincides with our DIA results of $U_c = 3.7$, where we observe a second order phase transition.  With increasing $\lambda_\tn{SO}$, all methods show that $U_c$ increases as well. Mean-field~\cite{rachel}, however, overestimates here the slope in comparison with the more elaborate methods~\cite{yu, budich, laubach, hohenadler1, hohenadler2, assaad, hohenadler}. The reason for that is analyzed in the Appendix~\ref{comp}. Our results show a similar behaviour as VCA with different cluster geometries~\cite{laubach}. To sum up, in the inversion-symmetric case the two-site DIA is in good agreement with other methods. We can therefore expect that the method is suitable to explore the model when inversion symmetry is broken.

\begin{figure}
 \centering
  \includegraphics[width=0.42\textwidth]{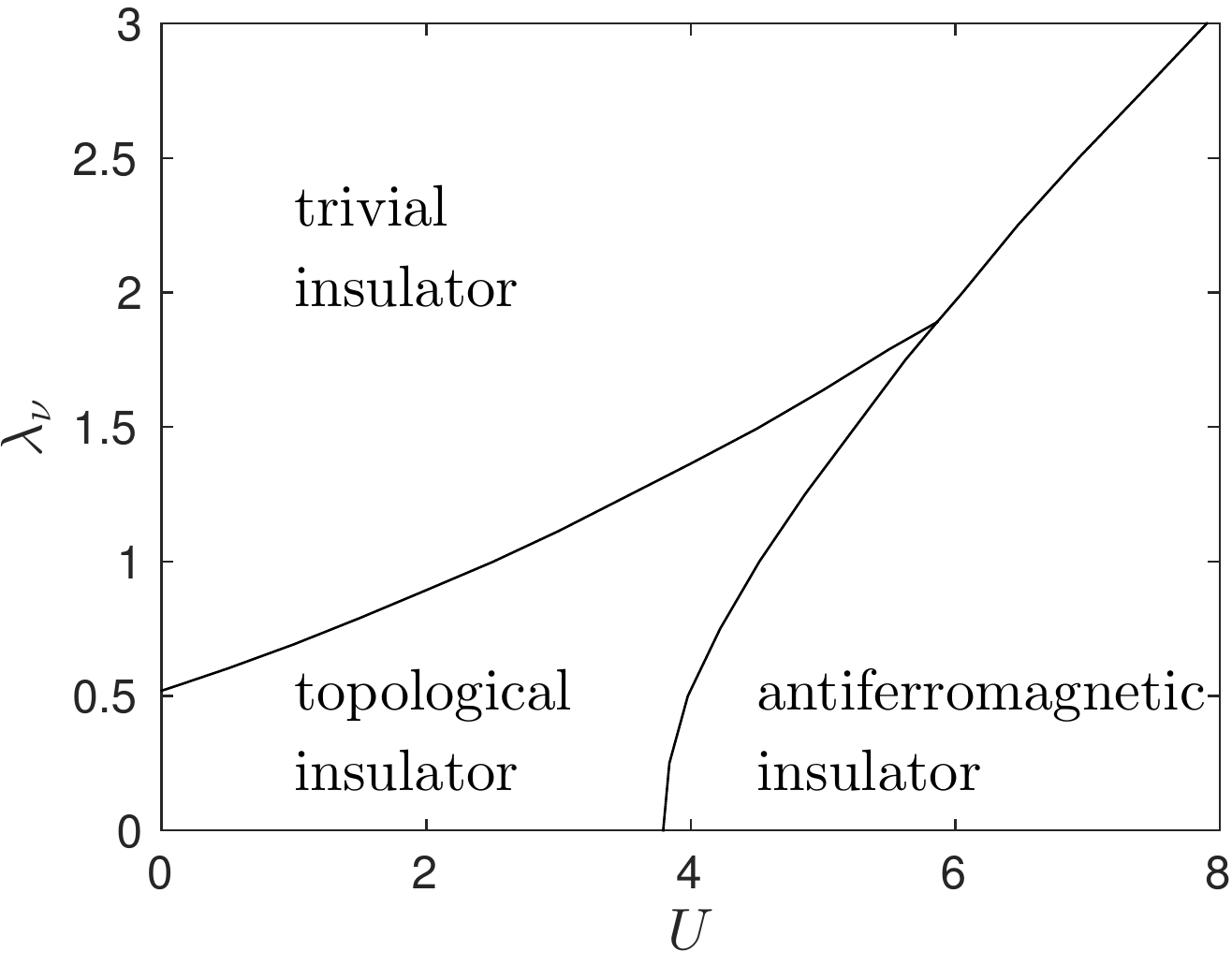}
 \caption{Phase diagram of the KMH model obtained from two-site DIA, as a function of the Hubbard interaction and the sublattice potential for a spin-orbit coupling of $\lambda_\tn{SO} = 0.1$.  \label{fig:5}}
\end{figure}

Using the topological Hamiltonian defined in Eq. (\ref{eq:Htop})
in combination with the Soluyanov-Vanderbilt method, information on
the topological properties can be obtained in addition to the magnetic
ordering. In the noninteracting case, a topological phase transition
occurs at $\lambda_\nu = 3 \sqrt{3} \lambda_\tn{SO}$, as known from
the original work by Kane and Mele~\cite{kanemele1,
  kanemele2}. Including a Hubbard interaction $U$, the topological
Hamiltonian has the same structure as the noninteracting Hamiltonian,
as long as the antiferromagnetic moment vanishes. However, both
self-energy and staggered on-site Weiss field renormalize the energy
scales. The interplay of interaction and on-site energy can be seen as
follows: Without interaction, the sublattice with the lower on-site
energy has a higher double occupancy. A finite Hubbard $U$ punishes
double occupancies, and reduces as a result the double occupancy on
the sublattice with lower on-site energy. Hence, the sublattice
potential $\lambda_\nu$ is effectively lowered in case of a finite
$U$, stabilizing the topological phase, and shifting the critical
$\lambda_\nu$ to higher values. The resulting phase diagram is shown
in Fig.~\ref{fig:5}. This stabilization effect is also captured in
mean-field, although with quantitative differences~\cite{lai}. 
We want to note that we cross-checked the validity of our WCC approach
by calculating
the spin Chern number $C_S$ directly from Eq.~(\ref{eq:cgreen}) for
the selected value of $U=1$. We
found perfect quantitative agreement.

\begin{figure}
  \includegraphics[width=0.37\textwidth]{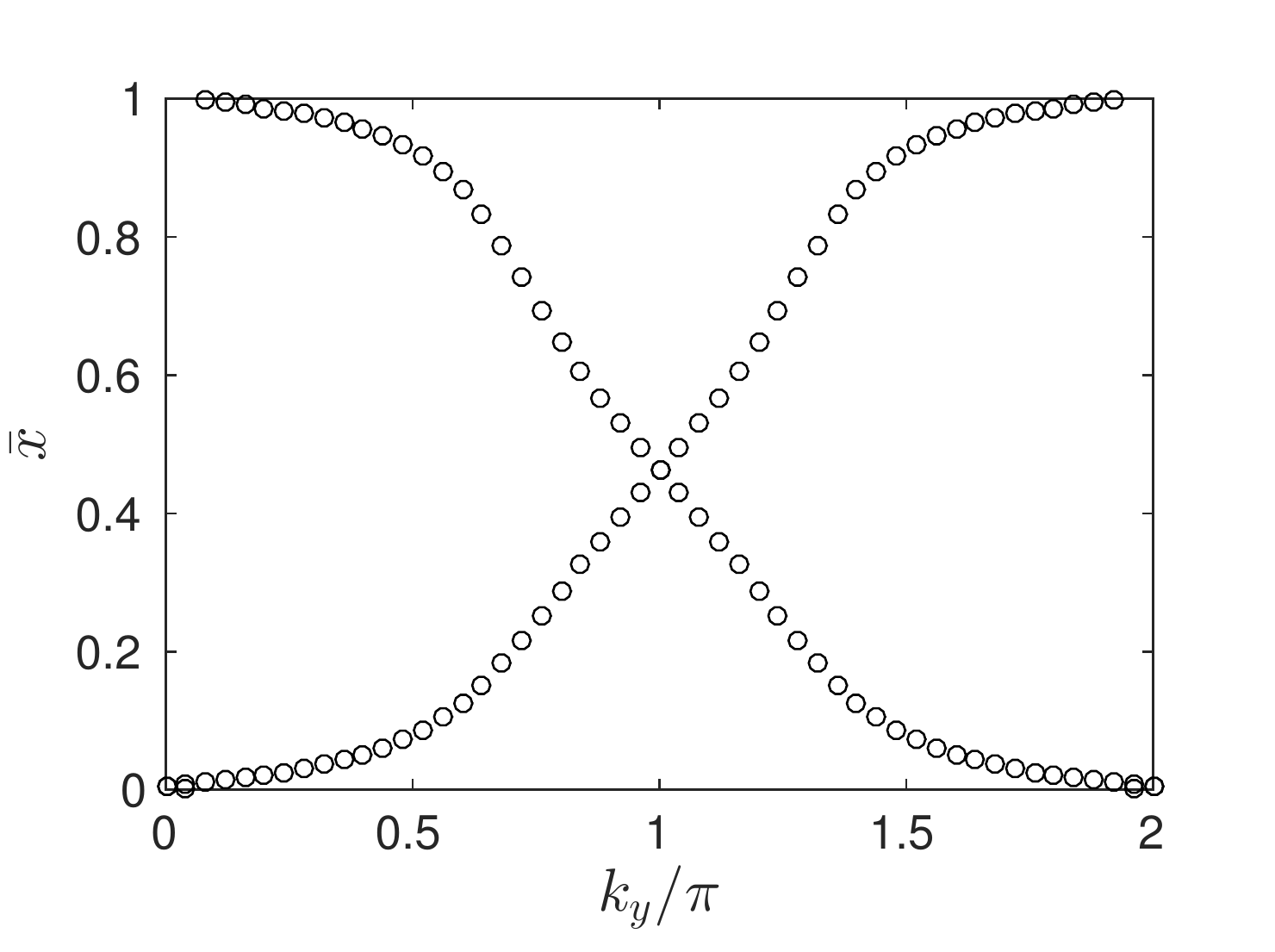}\\
  \includegraphics[width=0.37\textwidth]{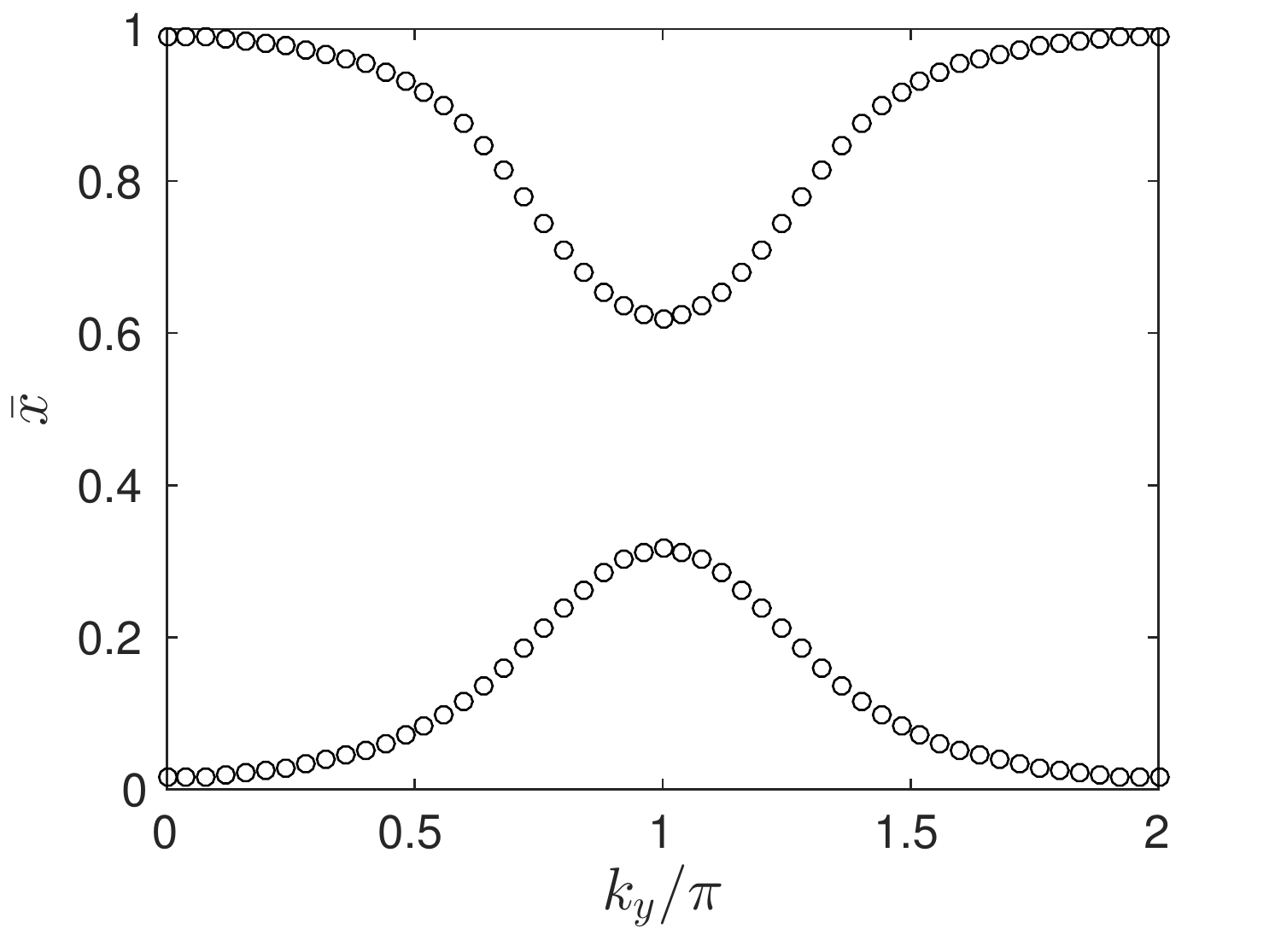}
 \caption{Wannier charge centers of the topological Hamiltonian for $\lambda_\tn{SO} = 0.1, \lambda_\nu = 0.25$, and $U = 3$ (top) and $U=4$ (bottom).\label{fig:6}}
\end{figure}

This reasoning for the stabilisation of the topological phase is only
valid in case of weak interactions where the antiferromagnetic Weiss
field is zero. In the strongly interacting regime, the non-vanishing
Weiss field causes a time-reversal symmetry breaking term proportional
to $\sigma_x \otimes \tau_z$ ($\sigma$ acts in spin space, $\tau$ in
sublattice space) in the topological Hamiltonian. As a consequence,
the topological invariant in the sense of Fu and Kane~\cite{fu2} is
not defined. This can also be seen in the WCC, where the lifted
Kramer's degeneracy does not enforce the two WCCs to be identical at
half the period of the pumping parameter. Examples of the WCCs are
shown in Fig.~\ref{fig:6}. In this regime, not just quantitative, but
also qualitative differences compared to a standard Hartree-Fock
mean-field arise, as discussed in Appendix~\ref{comp}. To sum up,
three phases exist for a given spin-orbit coupling: (i) a topological
insulator continuously connected to the quantum spin Hall phases of
the non-interacting KM-model if both $\lambda_\nu$ and  $U$ are small
enough; (ii) a trivial band insulator if $\lambda_\nu$ is large; (iii)
an antiferromagnetic insulator with in-plane magnetization for large
$U$. The phase boundaries are shown in
Fig.~\ref{fig:5}. Interestingly, similar results of an enhanced
topological phase have been reported for the Kane-Mele model
including long-ranged Coulomb
interactions~\cite{PhysRevB.90.085146}. There, the Coulomb interaction
induces charge-density-wave fluctuations, while our model shows static
charge ordering through staggered potentials. 

\subsection{Ribbon}

\begin{figure}
 \centering
  \includegraphics[width=0.45\textwidth]{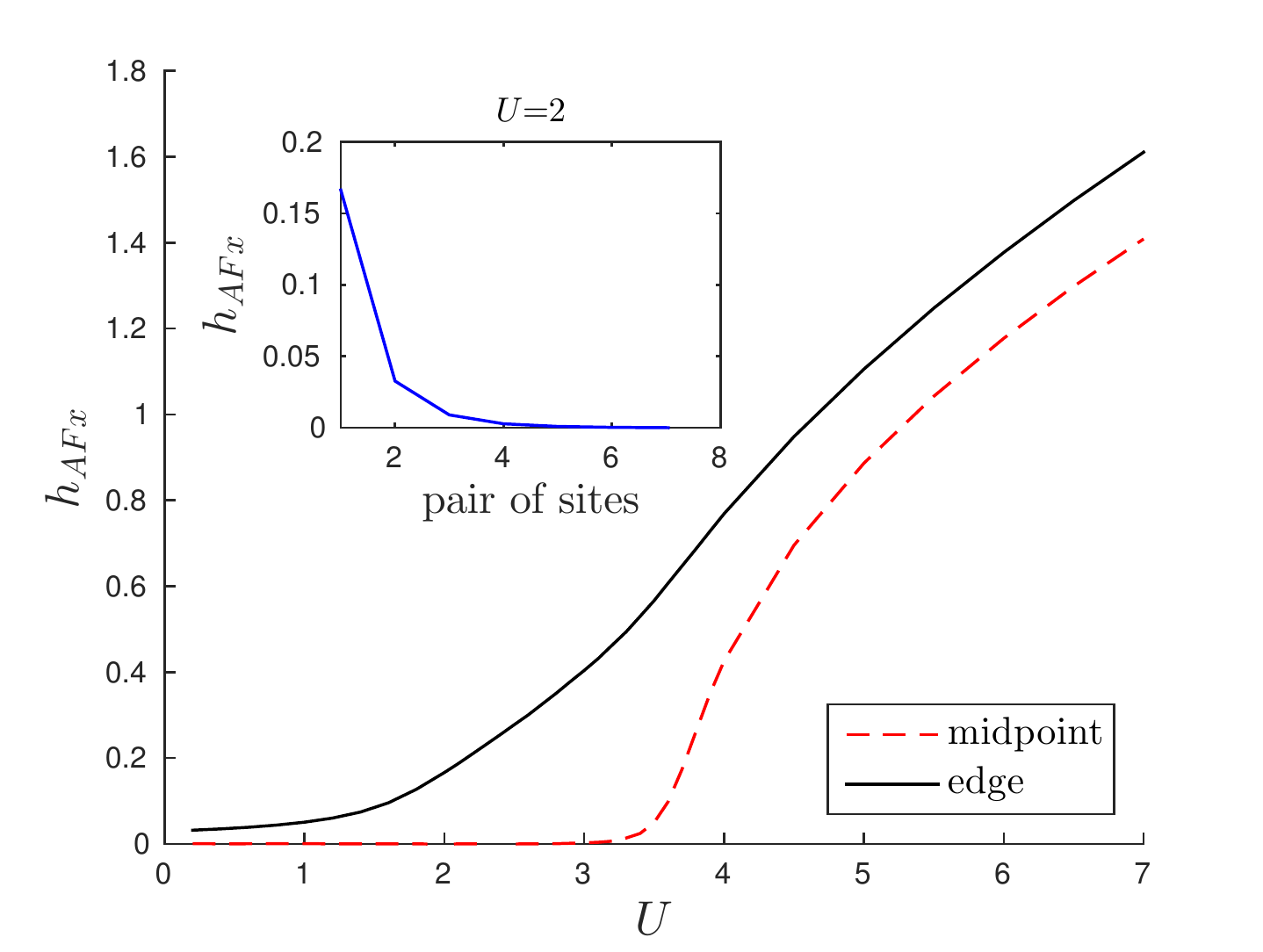}
 \caption{Antiferromagnetic Weiss field of the first pair of sites $A$ and $B$ of the ribbon (solid) and in the middle of the ribbon (dashed) as a function of $U$ for $\lambda_\tn{SO}=0.1$, $\lambda_\nu=0$ and $N = 16$ pairs of sites. The inset shows how the moment at the edge decays across the ribbon to the midpoint for $U=2$. \label{fig:7}}
\end{figure}

In order to analyze the robustness of the topological phases presented in the last section and to investigate the bulk-boundary correspondence, we calculate directly the edge properties on a zigzag ribbon of finite width. 

We first consider the inversion-symmetric case, $\lambda_\nu=0$. Mean-field results have shown different magnetizations at the edge than in the middle of the ribbon~\cite{lee}. This agrees with our results, and an example of the structure of the Weiss fields across the ribbon profile is shown in the inset of Fig.~\ref{fig:7}. The larger field at the edges decays quickly to the bulk value. The optimized values of both edge and midpoint antiferromagnetic fields as a function of $U$  are shown in Fig.~\ref{fig:7} for $\lambda_\tn{SO} = 0.1$. At the edges, any finite $U$ results in a finite antiferromagnetic field.
Sites that are not at the edges have a Weiss field comparable to the bulk values. Just below the bulk magnetic transition at $U\approx 3.8$ they become finite, though small, which is a finite-size effect caused by the increasing correlation length as the magnetic transition is approached. The main consequence of the non-vanishing Weiss field is that the finite magnetization at the edges breaks time-reversal symmetry for any $U$ and gaps therefore the edge states. 
As the interaction is below the critical value for the bulk magnetic
transition, topological analysis of the bulk suggests a topological
insulator with gapless edge states, but a local symmetry breaking at
the edges causes the edge states to gap. This local effect, namely
that local time-reversal symmetry breaking by a magnetic field causes
states to gap,  
cannot be captured within a topological invariant of the two-dimensional (2D)
system. However, at what point in the phase diagram this local symmetry
breaking occurs, depends both on the specific model and also on the
edge geometry. For example, for the armchair ribbon there is a region
at small $U$ with vanishing edge magnetization and therefore gapless
edge states, even in the
inversion-symmetric case $\lambda_\nu = 0$.  


In the last paragraph it is demonstrated that gapless edge states are impossible on a zigzag ribbon for any finite $U$, as long as $\lambda_\nu = 0$. This picture changes if inversion symmetry is broken. From the bulk calculations we know that $\lambda_\nu$ tends to suppress magnetic ordering, where it increases the critical value of interaction $U_c$ for the magnetic transition (Fig.~\ref{fig:5}). The same principle is observed looking at the edge magnetization as a function of $\lambda_\nu$. For given $U$ and $\lambda_\tn{SO}$, the Weiss field at the edges changes only marginally as $\lambda_\nu$ is increased, and the edge is magnetic. However, at a critical value $\lambda_\nu^c$, the magnetic moment drops to 0 in a first-order phase transition. This critical value $\lambda_\nu^c$ strongly depends on $U$. For $\lambda_\tn{SO} =0.1$, for example, we get $\lambda_\nu^c=0.006$ as $U=1$, and it raises by an order of magnitude to 
$\lambda_\nu^c=0.07$ for $U=2$ and to $\lambda_\nu^c=0.35$ for $U=3$. 

This argument can of course be turned around. Fixing the sublattice potential $\lambda_\nu$ and varying the interaction strength $U$, one finds a critical value $U_c$ for the magnetic transition with finite magnetic moment only for $U>U_c$. This critical value $U_c$ raises continuously with increasing sublattice potential $\lambda_\nu$, starting from $U_c=0$ at $\lambda_\nu=0$.

\begin{figure}
 \centering
  \includegraphics[width=0.37\textwidth]{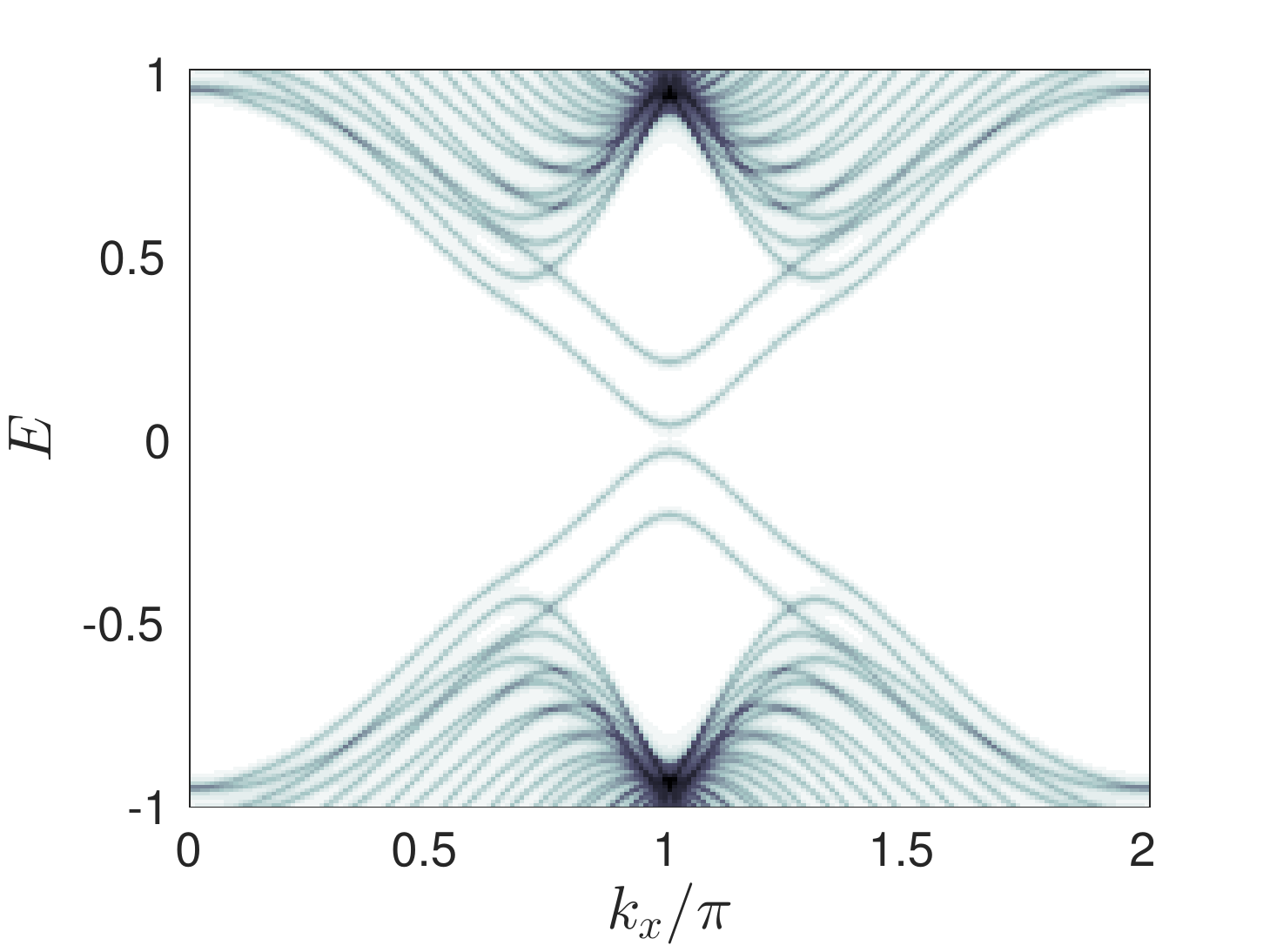}\\
  \includegraphics[width=0.37\textwidth]{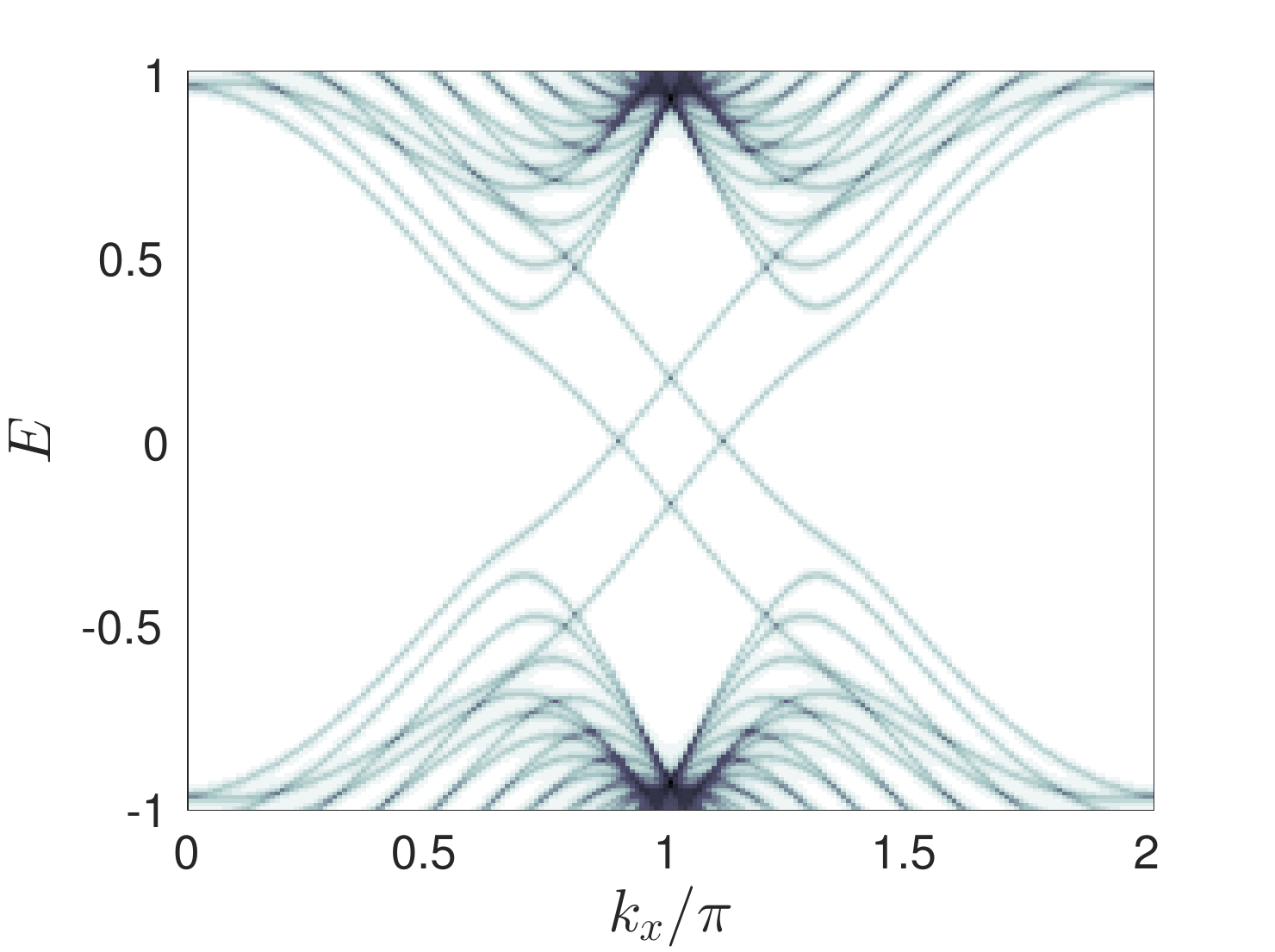}
 \caption{Spectral functions of the KMH zigzag ribbon with parameters $U=2.5$,  $\lambda_\tn{SO}=0.1$, and $N=16$. Top panel: $\lambda_\nu=0.1$ leads to a magnetic solution with a Weiss field of about $h_{\tn{AF}x} = 0.4$, gapping the edge states. Bottom panel: $\lambda_\nu=0.2$, with a vanishing Weiss field and gapless edge states. \label{fig:8}}
\end{figure}

Exemplary spectral functions are shown in Fig.~\ref{fig:8}, where we use spin-orbit coupling strength $\lambda_\tn{SO}=0.1$ and interaction strength $U=2.5$. If the sublattice potential $\lambda_\nu$ is below the critical value, as in the top panel of Fig.~\ref{fig:8}, the edge is magnetic and the edge states are gapped. For $\lambda_\nu>\lambda_\nu^c$ there is no magnetization at the edge, and gapless states occur. We want to stress again that gapless edge states do not occur at any finite $U$ in the inversion-symmetric case. To sum up, an inversion-symmetry-breaking term can stabilize the gapless edge state.

\section{Conclusion and Discussion}

We have investigated the topological properties of the Kane-Mele-Hubbard model, comparing cases with and without inversion symmetry. For the calculation of the topological invariants we apply a combination of the topological Hamiltonian approach and the Wannier charge center method. This approach allowed to calculate the phase diagram of the KMH model in the $U$-$\lambda_\nu$ plane. The inversion-symmetry-breaking term $\lambda_\nu$ has a two-fold effect. First, for large values the topological order is destroyed and a trivial insulator obtained. Second, in combination with interactions the topological order is enhanced, pushing the phase boundaries towards the antiferromagnetic insulator to larger critical values of $U$. 

This effect can also be seen in the surface properties of the
honeycomb lattice. In agreement with previous studies, our
calculations on the zigzag ribbon geometry have shown that with
inversion symmetry any finite value of $U$ results in a finite edge
magnetization, which in turn produces a finite gap in the edge
states. Introducing an inversion-symmetry-breaking field, this
critical value $U_c$ is shifted to finite values, below which the
whole ribbon including the edge is nonmagnetic, and a gapless surface
state exists. As a result, one can find gapless edge states on the
zigzag ribbon only when inversion symmetry is lifted and the
interaction strength $U$ is small enough, such that no ordered
magnetic moments can form.

Our study is based on the Kane-Mele Hamiltonian, which was introduced
as the low-energy Hamiltonian for graphene. Since the bulk gap in
graphene is minute, the effects that we propose here are difficult to
see in this material. However, there is increasing interest in
artificial honeycomb systems using heavy atoms, such as
bismuthene on SiC substrate~\cite{bismuthene}. Since these systems are
grown artificially, it might be possible to modify their structure
such that inversion symmetry is broken and the influence of this
symmetry breaking on the topological properties can be studied.

\acknowledgments
R.T. thanks Georg W. Winkler for helpful discussion. We acknowledge
financial support from the Austrian Science Fund FWF, START program
Y746. 

\appendix

\section{Discussion - Comparison to mean-field}\label{comp}

As mentioned in Sec.~\ref{bulk}, the basic structures of the topological Hamiltonian could also be found in a mean-field approximation since the self-energy is diagonal. Usually, the $z$ axis is chosen as the axis of mean-field decomposition~\cite{rachel}. The resulting matrix is then qualitatively different from the topological Hamiltonian of the DIA, since the mean-field magnetic moment points in the $z$ direction. In order to respect that the easy axis is in-plane, we did a mean-field decoupling in the $x$ direction 
\begin{equation} \label{eq:mfx}
 n_{i \uparrow} n_{i \downarrow}\approx
 \left( \langle n_{i \leftarrow}\rangle n_{i \rightarrow}+\langle n_{i \rightarrow}\rangle n_{i \leftarrow}-\langle n_{i \leftarrow}\rangle \langle n_{i \rightarrow}\rangle\right),
\end{equation}
where \renewcommand\arraystretch{0.2}$\ket{\begin{matrix} \rightarrow\\ \leftarrow \end{matrix}} = 1/\sqrt{2}\left(\ket{\uparrow} \pm \ket{\downarrow}\right)$. Within this framework, the same phases as in the DIA appear, where the mean-field one-electron Bloch Hamiltonian corresponds to the topological Hamiltonian. The phase boundaries, however, will shift since a bare mean-field approach does not capture quantum dynamics as the DIA.

\begin{figure}
 \centering
 \includegraphics[width=0.37\textwidth]{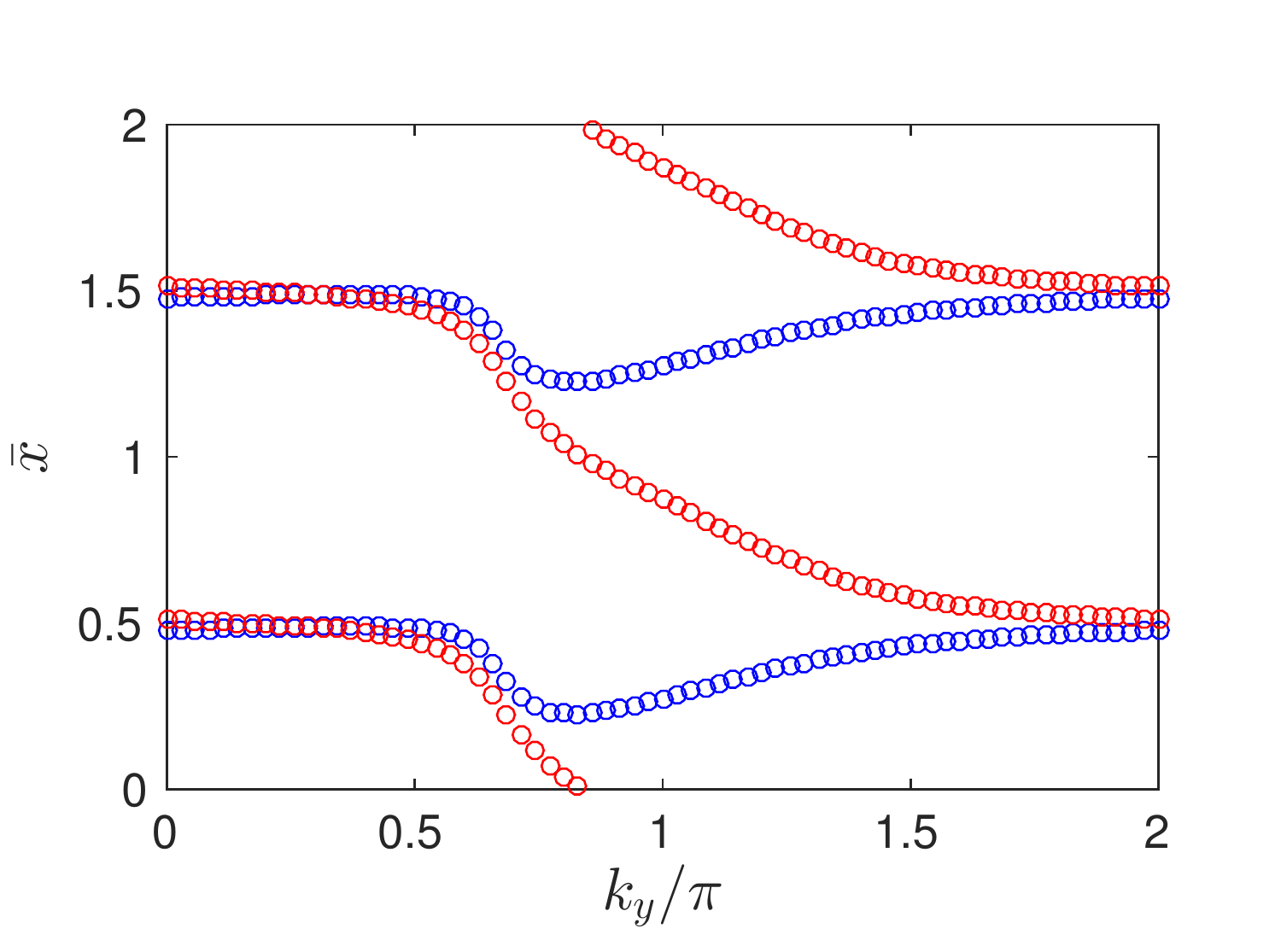}\\
 \includegraphics[width=0.37\textwidth]{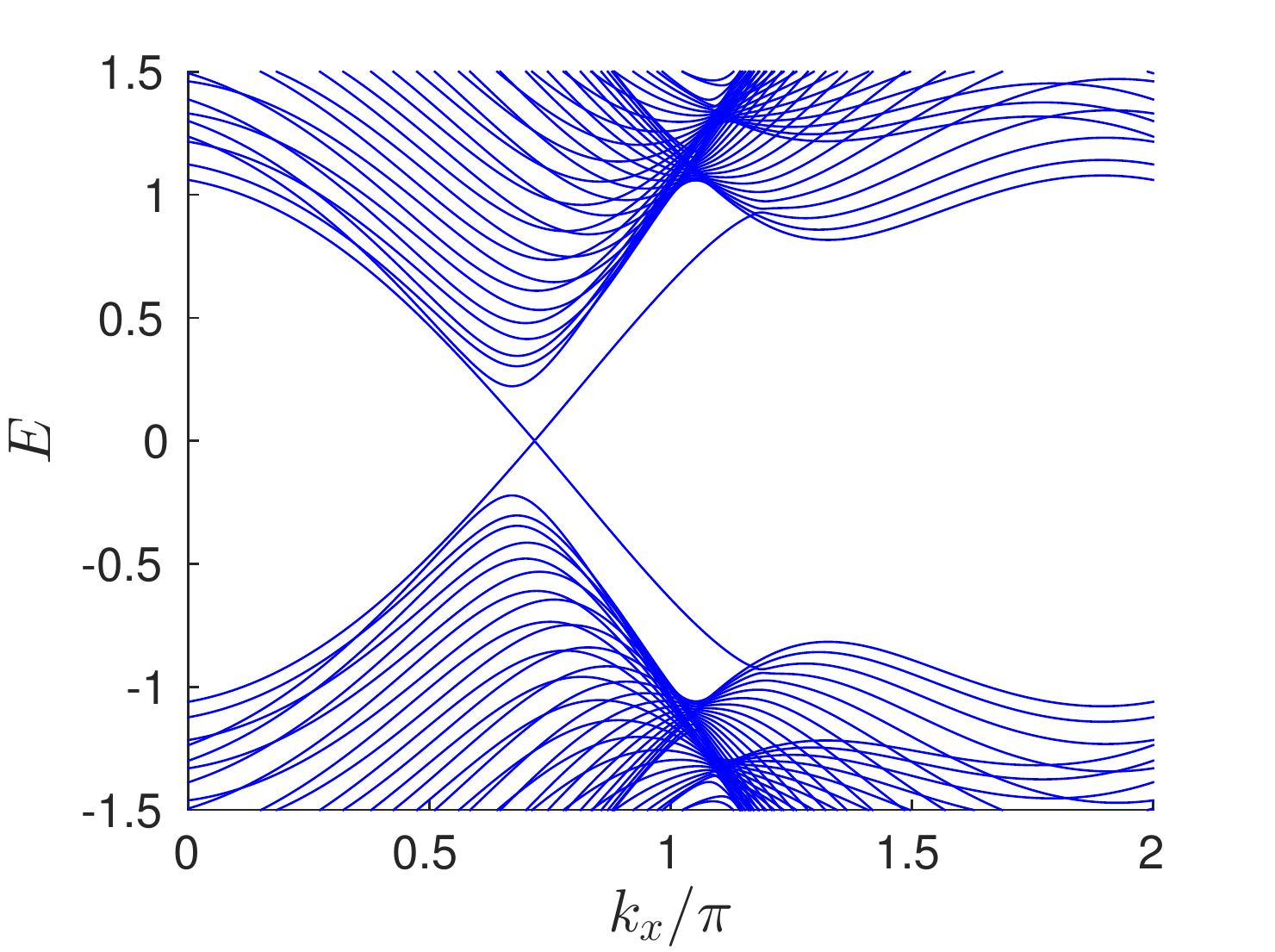}
 \caption{Chern insulator phase of the KMH model in mean-field approximation with antiferromagnetic moment in $z$ direction for $U=3.2$, $\lambda_\tn{SO}=0.1$, $\lambda_\nu=0.4$. The upper graph shows Wannier charge centers from the bulk calculations. The blue curve is the WCC of the spin up band, the red curve the WCC of the spin down band, resulting in $C\up = 0$ and $C\down = -1$. The lower graph shows the bands of a ribbon ($N=32$) with one spin down, but no spin up edge state. \label{fig:9}}
\end{figure}

In case of the Hubbard model on a honeycomb lattice $\lambda_\tn{SO}=\lambda_\nu=0$, the magnetization direction is not important since SU(2) symmetry is not broken. The mean-field critical interaction for any quantization axis is $U_c = 2.23$~\cite{rachel, sorella}. If $\lambda_\tn{SO}\neq 0$, the difference between the two mean-field methods is important. Since the in-plane magnetic moment is always favorable, a restriction of the magnetization direction to be out-of-plane requires stronger interactions for the stability of the antiferromagnetic solution. This is the case in a conventional mean-field theory~\cite{rachel,lai}, hence, $U_c$ is overestimated in comparison with an in-plane mean-field approach (\ref{eq:mfx}). Consequently, the slope of the $U_c$-$\lambda_\tn{SO}$ phase boundary is higher if $z$ is used as a quantization axis. 

In addition to the magnetic transition considered so far, using Wannier charge centers as an analytical tool allows again to extract topological information. The DIA results are described in the previous sections, showing the phase diagram of three different phases in figure \ref{fig:5}. As mentioned above, the mean-field decoupling in the $x$ direction gives qualitatively the same phases since the MF Bloch Hamiltonian has the same structure as the DIA topological Hamiltonian, but underestimates $U_c$. New phases appear, however, in the standard Hartree-Fock approach where the $z$ axis is the quantization direction. The Hamiltonian splits into spin up and spin down parts, which are decoupled if neither Rashba coupling nor in-plane magnetization are present. Hence, even though time-reversal symmetry is broken in the presence of an antiferromagnetic moment, a $\mathbb{Z}_2$ invariant can be defined using the spin Chern number $\nu_S = C_S\mod 2$, $C_S = (C\up - C \down)/2 $ as introduced by Sheng \textit{et al.}~\cite{sheng2}. The Chern numbers of the two spin categories are determined with the Wannier charge centers: Because of the conservation of $S_z$, the two WCC can be labeled by their spin. The Chern number $C_S$ is then given by the difference of the WCCs $\bar{x}\up$ and $\bar{x}\down$ as they evolve continuously from $0$ to $2\pi$. 

In the inversion-symmetric case, the only mean-field parameter that has to be determined self-consistently is the antiferromagnetic moment $M_\tn{AF}=\avg{n_{A \uparrow}} - \avg{n_{B \uparrow}} =\avg{n_{B \downarrow}} - \avg{n_{A \downarrow}}$. A change of both Chern numbers $C\up$ and $C\down$ occurs when the gap closes at a critical moment $M_\tn{AF}^c = 12\sqrt{3}/U$, which follows from diagonalizing the mean-field Bloch Hamiltonian. Since $M_\tn{AF}$ rises continuously from $0$ as $U$ is increased, magnetic and topological transition do not coincide, leading to an antiferromagnetic quantum spin Hall phase between the two transitions.

If additionally inversion symmetry is broken, both on-site energy
and occupation of $A$ and $B$ sites are different. Together with the
magnetic order, this leads to different $M_\tn{AF}^c$ for spin up and
spin down electrons. If $C\up = 0$ and $C\down =1$ or vice versa, the
total Chern number $C=C\up+C\down$ is nontrivial. Hence, for a certain
parameter range, an antiferromagnetic Chern insulator is
realized (see Fig. \ref{fig:9}). Both Chern insulator and antiferromagnetic quantum spin Hall
insulator have also been found recently for cases where the symmetry breaking is not due to an on-site potential, but due to a spin-dependent hopping \cite{miyakoshi}. These phases are stable since for certain parameter regions the out-of-plane magnetization is energetically favorable. 

The topological properties of the Chern insulator are not bound to time-reversal symmetry but related to the spin structure only. The number of edge states is directly determined by the Chern numbers of spin up and spin down electrons. As an example, the bands of a zigzag ribbon in the Chern insulator phase with only one edge state are shown in Fig. \ref{fig:9}. Hence, bulk boundary correspondence is fully satisfied if the antiferromagnetic moment is in the $z$ direction, but not if it is in-plane.

\bibliography{bib_paper.bib}

\end{document}